\documentclass[onecolumn,showpacs,showkeys,preprintnumbers,amsmath,amssymb]{revtex4}

\usepackage{graphicx}% Include figure files
\usepackage{dcolumn}% Align table columns on decimal point
\usepackage{bm}% bold math
\newcommand{\BE}{\begin{equation}}
\newcommand{\EE}{\end{equation}}
\newcommand{\BA}{\begin{eqnarray}}
\newcommand{\EA}{\end{eqnarray}}
\newcommand{\ND}{\noindent}

\begin{document}

%\preprint{}
%---------------------------

\title{
%Nonspreading wave packets in a general potential $V_{nswp}(x, t)$ in one dimension
%Nonspreading wave packets constructed from general potential $V_{nswp}(x, t)$
Analyzing and constructing general  nonspreading wave packets
% from general potential $V(x)$

}



\author{Chyi-Lung Lin}\email{cllin@scu.edu.tw}

%\author{Chyi-Lung Lin$^{1}$}
%email{cllin@scu.edu.tw}
%\author{Meng-Jie Huang$^{2}$}
%\author{Te-Chih Hsiung$^{3}$}

%\email[]
%\thanks{}

\affiliation{Department of Physics, Soochow University, Taipei,Taiwan, R.O.C. }

%\affiliation{$^{1}$Department of Physics, Soochow University, Taipei,Taiwan, R.O.C. \\
%$^{2}$Institut f\"ur Festk\"orperphysik,  Karlsruhe Institute of Technology, 76021 Karlsruhe, Germany \\
%$^{3}$Taiwan Semiconductor Manufacturing Company, Hsinchu Science Park, Hsinchu 300, Taiwan}

%\author{\thanks{ E-mail: \email{ }}}

%\institute{Department of Physics,\\ Soochow University, \\Taipei,Taiwan, %R.O.C.}
%\pacs{03.65.w}{ Quantum mechanics}


%\date{\today}
%------------------------------------

\begin{abstract}
{
We show the method for constructing nonspreading wave packets whose shape and motion can  be general.
We analyze the time evolution of nonspreading wave packets by decomposing the Hamiltonian into two parts. Of the two, one changes the instantaneous state, the other does not. Through this decomposition, the time evolution operator is shown to be effectively a spatial shifting operator. 
This explains why nonspreading wave packets can be nonspreading.
And we show that the part of the Hamiltonian which changes the instantaneous state governs the motion of the nonspreading wave packets. 
}

\end{abstract}




%\keywords{}

\maketitle


\section{\label{sec:INT}  Introduction}

Nonspreading wave packets (NSWPs) are  quantum packets whose probability density function, $|\Psi(x,t)|^2$, 
%the spatial distribution,
do not change their form while propagating.
There are stationary wave packets which correspond to  
the energy eigenstates of a Hamiltonian with a static potential energy.
These energy eigenstates may be viewed as stationary NSWPs, or the trivial NSWPs. These  stationary NSWPs are, in fact, important for constructing propagating NSWPs with shapes in a general form. 
%Intuitively, 
%it is possible to have nontrivial NSWPs, as if we look at a stationary energy eigenstate from an accelerated frame,  then we should view a propagating NSWP. 
%Since there are a variety of energy eigenstates, we should expect there are many other NSWPs besides those already known.  
We will discuss constructing general NSWPs in Section II.

The first nontrivial nonspreading wave packet was found by Schr\"{o}dinger \cite {Schrodinger} in 1926. 
This packet is normalizable and is of the form as a shifted ground state of a harmonic oscillator. 
The motion of the packet behaves just like a classical particle which is being shifted from its equilibrium position and executing a simple
harmonic motion. 
After Schr\"{o}dinger, Senitzky in 1954 extended the result showing
that shifted higher energy eigenstates are also nonspreading
\cite{Senitzky}. 
Senitzky also showed that the expectation value of the energy of a nonspreading wave packet is $E_n + E_{cl}$, where $E_n$
is the quantum mechanical energy of a wave packet when it is
stationary, and $E_{cl}$ the classical energy of the particle. 
Thus,
though the motion of a nonspreading wave packet is the same as a classical particle, the energy content of the packet is different from that of a classical particle. 
We may view
$E_n$ as the structural energy of a stationary wave packet besides its classical kinetic energy $E_{cl}$.
The extension of nonspreading packets from a shifted ground state to
a shifted higher energy eigenstates were also found by Roy and Singh
\cite{Roy},  Yan  \cite{Yan}, and Mentrup \textit{et al}.
\cite{Mentrup}, etc.

Much later than Schr\"{o}dinger, Berry and Balazs in 1979 found another
interesting nonspreading Airy packet in free space
\cite{Berry}. Airy packets, with a form described by Airy function $ \textmd{Ai}[x]$, though are bounded functions, however, are not square integrable. 
Especially, an Airy packet does not move in a uniform velocity
in free space; instead, it accelerates. 
Thus, 
in contrast to Schr\"{o}dinger's packets, the motion of an Airy wave packet is different from that of a classical particle.
This is due to that Airy packet is not normalizable, and therefore does not
really describe a particle. Nonspreading Airy packets even exist in time-dependent uniform force \cite{Berry}. 
%It had been shown that no other potentials except those described above will support nonspreading Airy packets \cite{Lin1}. 

There are a lot of
investigation on the interesting optical Airy beams. The Airy
optical beams was first observed by Christodoulides
\textit{el}. \cite{Christodoulides}. And electron Airy beams were also constructed by Noa \textit{el}.  \cite{Noa}
%
%
%Nonspreading wave packets in an imaginary potential was also observed by St\"{u}zle \textit{el}.\cite{Stuzle}.
%
%
The phenomenon of nonspreading wave packets is an interesting subject.
The essential ingredients of a nonspreading wave packet are: its
shape function $f(x)$, its motion $d(t)$, and its phase function
$\phi(x,t)$. 
%From the previous works of many people, we can have a general discussion on this subject. 
In what follows, we will show that stationary energy eigenstates can be used to generate propagating NSWPs with the same shapes.

In Section II, we discuss the method for constructing NSWPs from stationary states. In Section III, we explain why NSWPs can be nonspreading from the point of view of time evolution.
In Section IV, we use our method to reproduce Berry, Balazs, Schr\"{o}dinger and
Senitzky's NSWPs.
In section V, we generalize the formalism to three dimensions.

\section{\label{sec:II}
constructing NSWPs from arbitrary static potential energy function $\textbf{V(x)}$}

In this section, we discuss how to construct NSWPs from an arbitrary static potential energy function, $V(x)$.
%
% 
%In what follows, we discuss how to construct NSWPs with an arbitrary motion described by the function $d(t)$.
%
%
%We start from an initial wave packet described as follows
%\BE
%\Psi(x,0)=f(x)\;e^{i\theta(x)},  \label{2-1}
%\EE\\
%\ND where $f(x)$ and $\theta(x)$ are real functions.
The time evolved wave function $\Psi(x,t) $  of an NSWP with a trajectory described by $x=d(t)$ 
should be of the form as

\BA
&&
\Psi(x,t)=f(q)\;e^ {i
\phi(x,t)}, \label{2-1}
\\
&&
q= x-d(t),
\label{2-2}
\EA\\
wher $f(x)$ and $\phi(x,t)$ are  real functions, and $d(t)$ can be arbitrarily designed.
%, and $\phi(x,0) = \theta(x)$.
The function $d(t)$ describes the motion of the packet,
and $\dot{d}(t)$ is the group velocity of the packet.
We have 
%$|\Psi(x,0)|^2=f(x)^2$ and 
$|\Psi(x,t)|^2= |f(q)|^2$,
so that $|\Psi(x,t)|^2$
has the same form all the time.
%$|\Psi(x,0)|^2$;
%therefore, $\Psi(x,t)$ is a nonspreading wave packet.
%
%
%Thus, the function $f$ describes the shape of $|\Psi(x,t)|$. 
%
For the previous NSWPs obtained by Schr\"odinger and Senitzky, $f(x)$ is the eigenfunction of a Hamiltonian with $V(x)= (1/2) k x^2$. And $V(x)= A x$ for NSWPs obtained by Berry and Balazs; if    $A=0$, it means a free space.
%In fact, for a more general case,  $V(x)$ can be chosen arbitrarily.
%, and we start from this assumption. 
To construct NSWPs for a more general case, we let the shape function $f(x)$ be determined from the time independent Schr\"odinger equation
\BE
\Big[-\frac{\hbar^2}{2m}\partial_x\partial_x+V(x)\Big] f(x)= E_f f(x).
\label{2-3}
\EE\\
where $V(x)$ is an arbitrary static potential energy.

The next step is to determine a potential energy, denoted by $V_{nswp}(x,t)$, in which NSWPs described in Eqs.(\ref{2-1})-(\ref{2-2}) can be constructed. 
The time dependent Schr\"odinger equation for such potential energy is

\BE
i\hbar \partial_t \Psi(x,t)= H \Psi(x,t)
=
\Big[-\frac{\hbar^2}{2m}\partial_x\partial_x+V_{nswp}(x,t)\Big] \Psi(x,t),
\label{2-4}
\EE\\
%
%
%We easily obtain
%
%
%\BE
%\partial_x \partial_x \Psi(x,t)=f''(q) e^{i \phi(x,t)} +2 i \; \partial_x \phi(x,t) \; \partial_x \Psi(x,t)+(\partial_x \phi(x,t))^2
%\label{2-6}
%\EE\\
%
%
We can determine $V_{nswp}(x,t)$ from  Eq.(\ref{2-4}).
We first rewrite Eq.(\ref{2-4}) as

\BE
i\hbar \partial_t \Psi(x,t)
+
\frac{\hbar^2}{2m}\partial_x\partial_x \Psi(x,t)
=
V_{nswp}(x,t)\Psi(x,t),
\label{2-5}
\EE\\
Substituting the $\Psi(x,t)$ in Eqs.(\ref{2-1})-(\ref{2-2}) to the left side of Eq.(\ref{2-5})
and comparing it with the right side, we can then determine the $V_{nswp}(x,t)$.
For the first term of the left side of Eq.(\ref{2-5}), we  have

\BE
i \hbar  \partial_t \Psi(x,t)
=
\Big[
\dot{d}(t) P 
-
\dot{d}(t) \;  
\hbar \partial_x \phi(x,t)
-
\hbar \partial_t \phi(x,t)
\Big]
\Psi(x,t).
\label{2-6}
\EE\\ 
where, $P= -i \hbar \partial/ \partial x$ is the momentum operator. For the second term  of the left side of Eq.(\ref{2-5}), we have

\BE
\partial_x \partial_x \Psi(x,t)=e^{i \phi(x,t)} \partial_q \partial_q f(q) 
+2 i \; \partial_x \phi(x,t) \; \partial_x \Psi(x,t)+\Big( \partial_x \phi(x,t) \Big)^2 \Psi(x,t).
\label{2-7}
\EE\\
The term $\partial_q \partial_q f(q)$ above can be replaced by $V(q)$ and $E_f$. This is due to that, from Eq.(\ref{2-3}), we have the following formula
% for $\partial_q \partial_q f(q)$
%
%
\BE
\Big[-\frac{\hbar^2}{2m}\partial_q\partial_q+V(q)\Big] f(q)= E_f f(q).
\label{2-8}
\EE\\
Substituting Eq.(\ref{2-8}) into Eq.(\ref{2-7}), we obtain
\BE
\frac{\hbar^2}{2m}
\partial_x \partial_x  \Psi(x,t)
=
\Big[
V(q)
-
E_f
-
\frac{\hbar}{m} \partial_x \phi(x,t)\;  P 
+
\frac{\hbar^2}{2m} ( \partial_x \phi(x,t) )^2 
\Big]
\Psi(x,t).
\label{2-9}
\EE\\ 
Adding Eqs.(\ref{2-6}) and (\ref{2-9}) and comparing it with Eqs.(\ref{2-5}), we then obtain the form of $V_{nswp}(x,t)$. We have
\BE
V_{nswp}(x,t)
=
\Big[  
\dot{d}(t)-\frac{\hbar}{m} 
\partial_x \phi(x,t)
\Big] P
-\dot{d}(t) \hbar \partial_x \phi(x,t)
- \hbar \partial_t \phi(x,t)
+
V(q)
-E_f
+\frac{\hbar^2}{2m} (\partial_x \phi(x,t))^2.
\label{2-10}
\EE
%
%
%We are to construct a time dependent Schr\"odinger equation with a potential energy $V(x,t)$ from the subtraction of Eqs. (\ref{2-6}) and (\ref{2-7}).
%
%
This is the general form of $V_{nswp}(x,t)$.
We consider this general form elsewhere for the case that EM interactions are involved. 
At present, we do not consider a potential energy that is $P$ dependent. 
Then, the term containing $P$ in the right side of Eq.(\ref{2-10}) should be zero. This then yields
\BE
\dot{d}(t)
=
\frac{\hbar}{m} \partial_x \phi(x,t).
\label{2-11} 
\EE\\ 
From Eq.(\ref{2-11}), $\phi(x,t)$ can be easily determined, we have 
\BA
\phi(x,t)
&=& \phi_1(t) x +\phi_o(t).
\label{2-12}
\\
\phi_1(t) 
&=& 
\frac{ m \dot{d}(t)}{\hbar}.
\label{2-13}
\EA\\
Where $\phi_0(t)$ is an arbitrary  function depending only on time. We see that $\phi(x,t)$ is simply linear in $x$. We previous had also obtained this result in another way \cite{Lin1}.  
%The results of Eqs.(\ref{2-9})-(\ref{2-10}) can be seen from the known solutions of NSWPs \cite{Schrodinger, Senitzky, Berry} 
%The detail of the idea in obtaining  $V_{nswp}(x, t)$ is referred to Ref.\cite{Lin2}.
%
%
Substituting Eqs. (\ref{2-12})-(\ref{2-13}) into Eq. (\ref{2-10}), we obtain 

\BA
&&
V_{nswp}(x,t)
=
V(q)
-m \ddot{d}(t) x 
+G(t),
\label{2-14}
\\
&&
G(t)
\equiv
-E_f
-\frac{m}{2} \dot{d} (t)^2
- \hbar \dot{\phi}_0(t).
\label{2-15}
\EA
We see that $V_{nswp}(x,t)$ contains the original potential energy $V(x)$ but with the argument $x$ replaced by $q$. 
The function $G(t)$ is a function depending only on time. 
Since $G(t)$ depends only on time, it has no influence on dynamics. We may simply set $G(t)=0$. However, it is better to keep this term for the sake of later convenience. Once we have chosen a functional form for $G(t)$, we can then determine $\phi_0(t)$ from eq.(\ref{2-15}). The result is shown in Eq.(\ref{2-22}). 
%In general, the choice of the value for $G(t)$ is from the consideration that $\phi_0(t)$ can be simpler.
%
%

We thus have achieved the determination of the potential energy $V_{nswp}(x,t)$
that supports the NSWPs.
We conclude our results as the following:
\BA
&&
i\hbar\frac{\partial \Psi(x,t)}{\partial t}= H\Psi(x,t).
\label{2-16}
\\
&&
H =-\frac{\hbar^2}{2m}\partial_x\partial_x+V_{nswp}(x, t).
\label{2-17}
\\
&&
V_{nswp}(x,t)=V(q)-m \ddot{d}(t) x + G(t).
\label{2-18}
\EA

This Hamiltonian system, Eqs.(\ref{2-16})-(\ref{2-18}), supports NSWPs for which the wave packets are of the form as:  
\BA
&&
\Psi(x,t)=f(q)\;e^{ i \phi(x,t)}, \;\;\;\;\;\;\;q=x-d(t). 
\label{2-19}
\\
&&
\Big[-\frac{\hbar^2}{2m}\partial_x\partial_x+V(x)\Big] f(x)= E_f f(x).
\label{2-20}
\\
&&
\phi(x,t)=
\frac{m \dot{d}(t)}{\hbar} x +  \phi_0(t),
\label{2-21}
\\
&&
\phi_0(t)=
-
\frac{1}{\hbar}
\int_{0}^{t} 
\Big(E_f + G(t)+\frac{m}{2} \dot{d}(t)^2 \Big) dt.
\label{2-22}
\EA\\
Where we simply set $\phi_0(0)=0$. We easily check that the $\Psi(x,t)$
given in Eqs.(\ref{2-19})-(\ref{2-22}) satisfies the Schr\"odinger equation described in Eqs.(\ref{2-16})-(\ref{2-18}).
%We then have constructed NSWPs from an arbitrary potential energy $V(x)$.
This then completes the construction of NSWPs.
We start from an arbitrary static  potential function $V(x)$, then we obtain the potential energy $V_{nswp}(x,t)$ which supports NSWPs.

From Eq. (\ref{2-18}), we note that there is a moving potential energy $V(q)$ which moves along with the packet. This seems reasonable, if we are having a propagating nonspreading wave packet. 
There is also a potential energy term $- m x \ddot{d}(t)$. This term can be understood from the classical mechanics. As, classically, a particle with a  trajectory $d(t)$, then the force needed for that motion is $ F= m \ddot{d}(t)$, which corresponds to a potential energy term as  $- m \ddot{d}(t) x$.
%Since a quantum packet is just like a classical particle, the Hamiltonian therefore also contains a  potential energy term,$- m \ddot{d}(t) x$.

%, and there is also a force term $m \ddot{d}(t)$.  This force term can be understood from the point of view of classical mechanics. to achieve this motion, as classically the force term needed is $m \ddot{d}(t)$ which is offered by the potential energy term $- m x \ddot{d}(t)$. 

%
%
As an example for what we have constructed, we consider a time independent potential such as

\BE 
V(x)= \lambda x^4.
\label{2-23} 
\EE\\
The eigenfunctions, $f_{\lambda}(x)$, obtained from this potential energy offer the shape of NSWPs.
The NSWP-constructing potential is:
\BE 
V_{nswp}(x,t)= \lambda (x-d(t))^4 - m x \ddot{d}(t) + G(t),
\label{2-24} 
\EE\\
\ND 
where $d(t)$ can be arbitrarily designed. This potential energy $V_{nswp}(x,t)$ then provides NSWPs with their shapes being of the form as $f_{\lambda}(x)$.
The wave functions $\Psi(x,t)$, described in  Eqs.(\ref{2-19})-(\ref{2-22}), then are the NSWPs moving with a trajectory $x=d(t)$. 
%We note that there is a moving potential energy , just as the moving potential does. 

\section{\label{sec:III}
the time evolution and the dynamics of NSWPs }
 
In this section, we discuss how nonspreading wave packets can be nonspreading.
We start from the time evolution operator.
The infinitesimal time evolution operator is defined as:
\BA
&&
\Psi(x,t+dt)= U(t,t+dt) \Psi(x,t),
\label{3-1}
\\
&&
U(t,t+dt) = e^{-\frac{i}{\hbar} H(t) dt}.
\label{3-2}
\EA
We investigate how the wave function $\Psi(x,t)$ changes with time. For that, we decompose the Hamiltonian $H(t)$ into two parts, $\tilde{H}(t)$ and $H_c (t)$. The first part, $\tilde{H}(t)$, is the one that does not change the instantaneous state $\Psi(x,t)$. The second part, $H_c (t)$, does change  $\Psi(x,t)$.
To determine $\tilde{H}(t)$, we substitute  Eqs.(\ref{2-11})-(\ref{2-13}) into Eq.(\ref{2-9}). This yields the following interesting equation
\BE
\Big[
-\frac{\hbar^2}{2m}
\partial_x \partial_x +V(q)
-\dot{d}(t) P
\Big]
\Psi(x,t)
=
\Big[
E_f
-
\frac{m}{2} \dot{d}(t)^2
\Big]
\Psi(x,t).
\label{3-3}
\EE
This is an eigenvalue equation for  $\Psi(x,t)$. We  write the left side in terms of $\tilde{H}(t)$, and then the equation is written as 
\BA
&&
\tilde{H}(t) \Psi(x,t)
=
\tilde{E}(t) \Psi(x,t),
\label{3-4}
\\
&&
\tilde{H(t)}
=
-\frac{\hbar^2}{2m}
\partial_x \partial_x +V(q)
-\dot{d}(t) P,
\label{3-5}
\\
&&
\tilde{E}(t)
=
E_f
-
\frac{m}{2} \dot{d}(t)^2.
\label{3-6}
\EA
Thus the Hamiltonian $\tilde{H}(t)$ in  Eq. (\ref{3-5}) is the Hamiltonian that we are looking, as $\tilde{H}(t)$ does not change the state $\Psi(x,t)$. 
We then decompose the Hamiltonian  $H(t)$ into the form as: $H= \tilde{H}(t)+ H_c(t)$. Then  $H_c(t) = H(t)-\tilde{H}(t)$. This then determines  $H_c(t)$ which  is the part of the Hamiltonian that changes the state $\Psi(x,t)$. We have therefore obtained the results:
\BA
&&
H(t)=\tilde{H}(t)+ H_c(t)
\label{3-7}
\\
&&
H_c(t)=\dot{d}(t) P- m \ddot{d}(t) x +G(t)
\label{3-8}
\EA
With this form, then
\BA
&&
U(t,t+dt)  \Psi(x,t)
\label{3-9}
\\
&&
=
 e^{-\frac{i}{\hbar} H_c(t)dt} 
 e^{-\frac{i}{\hbar}\tilde{H}(t)dt} \; 
\Psi(x,t)
\label{3-10}
\\
&&
=
e^{-\frac{i}{\hbar}(\tilde{E}(t)+G(t))dt} \;  
e^{\frac{i}{\hbar} m \ddot{d}(t) dt x}
\;
e^{-\frac{i}{\hbar}\dot{d}(t) dt P}
\Psi(x,t)
\label{3-11}
\EA
We see that the time evolution operator $U(t,t+dt)$ is effectively a spatial shifting operator: $e^{-i \dot{d}(t) dt P/\hbar}$. It is well known that $e^{i a(t) P/ \hbar}$ shifts a spatial function $h(x)$ into $h(x+a(t))$.
Thus, we see that the operation of the  infinitesimal operator $U(t, t+dt)$ acting 
on the state $\Psi(x,t)$ transfers the state into a  shifted state $\Psi(x+\dot{d}(t) dt ,t))$
%
%
% that shifts $\Psi(x,t)$ to $\Psi(x+d(t) dt,t)$.
%$e^{-i\dot{d}(t) dt P/ \hbar}$. 
%
%
This explains why nonspreading packet can be nonspreading. 
The amount of the spatial shift is $dx= \dot{d}(t) dt$. Thus, $\dot{d}(t)$ is the group velocity of the NSWP.

We now discuss formally the operation of $U(t,t+dt)$ on the state $\Psi(x,t)$. 
From Eq.(\ref{3-11}), we need first calculate the operation of 
$e^{-\frac{i}{\hbar}\dot{d}(t) dt P}$ acting on $\Psi(x,t)$, and then the operation of $e^{\frac{i}{\hbar} m \ddot{d}(t) dt x}$, and then the operation of $e^{-\frac{i}{\hbar}(\tilde{E}(t)+G(t))dt} $.
For the operation of $e^{-\frac{i}{\hbar}\dot{d}(t) dt P}$ acting on $\Psi(x,t)$, we have  
\BA
e^{-\frac{i}{\hbar}\dot{d}(t) dt P} \Psi(x,t)
&=&
e^{-\frac{i}{\hbar}\dot{d}(t) dt P} f(x-d(t))\
e^{i (\phi_1(t) x + \phi_0(t))},
\nonumber
\\
&=&
f ( x-d(t)-\dot{d}(t) dt) 
\ e^{i (\phi_1(t) x - \frac{m}{\hbar} \dot{d}(t)^2 dt  + \phi_0(t))},
\nonumber
\\
&=&
f ( x-d(t+dt) ) 
\ e^{i (\phi_1(t) x  - \frac{m}{\hbar} \dot{d}(t)^2 dt + \phi_0(t))}.
\label{3-12} 
\EA
Above, we have used Eq.(\ref{2-13}). We see that under the operation of $e^{-\frac{i}{\hbar}\dot{d}(t) dt P}$, the original 
$f(x-d(t))$ is shifted to $f(x-d(t+dt))$,
and bring in an extra phase factor $ e^{-i \frac{m}{\hbar} \dot{d}(t)^2 dt}$.
Next, we discuss the operation of the phase factor   $e^{i m \ddot{d}(t) dt x/\hbar}$ acting on both sides of Eq.(\ref{3-12}). From the right side of Eq.(\ref{3-12}), we see that this phase factor $e^{i m \ddot{d}(t) dt x/\hbar}$ can be absorbed to the phase factor $ e^{i \phi_1(t) x}$. As  

\BA
e^{\frac{i}{\hbar} m \ddot{d}(t) dt x} 
e^{i \phi_1(t) x}
& =&
e^{\frac{i}{\hbar} m ( \dot{d}(t) + \ddot{d}(t)  dt) x}
\nonumber
\\
&=&
e^{\frac{i}{\hbar} m  \dot{d}(t+dt) x}
\nonumber
\\
&=&
e^{i \phi_1(t+dt) x}
\label{3-13}
\EA
We then have the following result obtained form the two successive  operations:
\BE
e^{\frac{i}{\hbar} m \ddot{d}(t) dt x} 
e^{-\frac{i}{\hbar} \dot{d}(t) dt P} \Psi(x,t)
=
f ( x-d(t+dt) ) 
\ e^{i (\phi_1(t+dt) x  - \frac{m}{\hbar} \dot{d}(t)^2 dt  + \phi_0(t))}
\label{3-14}
\EE
Finally, we discuss the operation of the phase factor 
$e^{-i(\tilde{E}(t)+G(t))dt/ \hbar}$ acting on both sides of Eq. (\ref{3-14}).
We also easily see that this phase factor 
can be absorbed into the phase factor $ e^{i \phi_0(t)}$.
As, from the right side of  Eq. (\ref{3-14}), we see we need calculate the following formula: 
\BA
e^{-\frac{i}{\hbar}(\tilde{E}(t)+G(t))dt }
e^{i ( - \frac{m}{\hbar} \dot{d}(t)^2 dt  + \phi_0(t))}
& = &
e^{-\frac{i}{\hbar}(\tilde{E}(t)+G(t) 
+ m \dot{d}(t)^2) dt} e^{i \phi_0(t)}
\nonumber
\\
&=&
e^{-\frac{i}{\hbar}(E_f +G(t) 
+ \frac{m}{2} \dot{d}(t)^2) dt} 
e^{i \phi_0(t)}
\nonumber
\\
&=&
e^{i(\phi_0(t)
+ \dot{\phi_0}(t) dt)}
\nonumber
\\
&=&
e^{i \phi_0(t+dt)}
\label{3-15}
\EA
Above, we have used Eqs. (\ref{2-22}), (\ref{3-6}).
In all, from all of these successive operations, we have then obtained that
\BA
\Psi(x,t+dt)
&=&
U(t,t+dt) \Psi(x,t)
\nonumber
\\
&=&
e^{-\frac{i}{\hbar}(\tilde{E}(t)+G(t))dt} \;  
e^{\frac{i}{\hbar} m \ddot{d}(t) dt x}
\;
e^{-\frac{i}{\hbar}\dot{d}(t) dt P}
 f( x-d(t) )
 e^{i (\phi_1(t)x + \phi_0(t) )}
\label{3-16}
\\
&=&
f( x-d(t+dt))
 e^{i (\phi_1(t+dt)x + \phi_0(t+dt))}
\label{3-17}
\EA
%
%
%\section{\label{sec:IV} the dynamics}
%
%
The action of the infinitesimal time evolution operator $U(t,t+dt)$ on the state $\Psi(x,t)$ indeed  remains the functional form of an NSWP.

As NSWPs move like classical particles, the dynamics of the motion should then be governed by an effective  Hamiltonian. From Eq.(\ref{3-7}), we see that the Hamiltonian is decomposed into two parts. The part of the Hamiltonian, $H_c(t)$, is the part of the Hamiltonian which involves the action of shifting the state $\Psi(x,t)$. 
The motion of NSWPs is thus effectively determined by $H_c(t)$ only. And, therefore, if we take an NSWP as a classical particle, then 
we expect that $H_c(t)$ would play classically the same role as a Hamiltonian that governs the motion of a particle. To verify that, we check the Hamilton equations obtained from the Hamiltonian $H_c(t)$. We have 

\BA
&&
\dot{x} =  \frac{\partial H_c}{\partial P}= \dot{d}(t),
\label{3-18}
\\
&&
\dot{P} =  \frac{\partial H_c}{\partial x}= m \ddot{d}(t).
\label{3-19}
\EA
This shows that indeed the Hamiltonian $H_c(t)$ governs the motion of NSWPs.
We should note that it is $H_c(t)$, not the total Hamiltonian $H(t)$, that governs the motion of NSWPs. This then explains why NSWPs can be accelerated in free space, as the corresponding  $H_c(t)$ in the case of a free space is not a constant, it is linear in $x$ and therefore accelerates a wave packet.  We discuss this in more detail in Section IV. 

In general, the function $V_{nswp}(x,t)$ is time dependent. 
It would be interesting to investigate whether it is possible for static potential energy systems to supply NSWPs. In fact, this is possible, and these potential energies are just those explored by Schr\"odinger and Berry and Balaz.
Below, we show that.

\section{\label{sec:IV} Deriving previous known NSWPs }

\subsection{\label{subsection:IV-1} Berry and Balazs's NSWPs (1) }

%\textbf{ 4-1. Berry and Balazs's NSWPs (1)}

In what follows, we show that we can derive the known examples of NSWPs from the method introduced in Sec. II.   
For the first case, we consider: $V(x)= A x$, where A is a real constant. This potential energy leads to the eigenvalue equation:
\BE
-\frac{\hbar^2}{2m}\partial_x\partial_x f(x) + A x f(x) = E_f f(x).
\label{4a-1}
\EE
The solution of $f(x)$ is expressed in terms of Airy function $Ai(x)$ as the following
\BE
f(x)= Ai\Big(\frac{2 A m }{\hbar^2}\Big)^{1/3}
\Big( x-\frac{E_f}{A} \Big).
\label{4a-2}
\EE
$f(x)$ will then be used as the shape function of the NSWPs.
The potential energy $V_{nswp}(x,t)$ which provides  NSWPs can be obtained from (\ref{2-18}). We have 
\BA
V_{nswp}(x,t)
&=&
A\;(x-d(t)) - m \ddot{d}(t) x +G(t)
\label{4a-3}
\\
&=&
\Big( A - m \ddot{d}(t) \Big) x + \Big( G(t)- A d(t) \Big),
\label{4a-4}
\EA
where $d(t)$ describes the motion of the packet.
If we consider the case that $V_{nswp}(x,t)$ is time independent, then 
the values for $A$ and $G(t)$ can be chosen as follows:
%If we are interested that $V_{nswp}(x,t)$ being a static potential, then we need
%
%
\BA
&&
A - m \ddot{d}(t)=0,
\label{4a-5}
\\
&&
G(t)- A d(t)=0.
\label{4a-6}
\EA

\ND
This then means that
\BE
V_{nswp}(x,t) =0.
\label{4a-7}
\EE\

\ND
Eq. (\ref{4a-7}) describes a free space. We thus have the interesting result: a free space system can offer NSWPs. This was found by Berry and Balazs in 1979.
Eq. (\ref{4a-5}) shows that the motion is with a constant acceleration, as we have $d(t)= (1/2) (A/m) t^2$, where we set $d(0)=0$. And then we have $G(t)= A^2 t^2/(2m)$ from Eq.(\ref{4a-6}).
From Eqs.(\ref{2-21})-(\ref{2-22}), we also obtain

\BA
\phi(x,t)
& = &
A t x +\phi_0(t),
\nonumber
\\
\phi_0(t)
& = &
-
\Big(E_f\; t +
\frac{A^2 t^3 }{3 m}  \Big),
\label{4a-8}
\EA\\
\ND
where we set $\phi_0(0)=0$.
From Eq. (\ref{2-19}), the full wave function of the NSWP is therefore

\BE
\Psi(x,t)
=
Ai\Big[
\Big( \frac{2 A m }{\hbar^2} \Big)^{1/3}
\Big(x-\frac{E_f}{A}-\frac{A t^2}{2m} \Big)
\Big]\;
e^{\frac{i}{\hbar} \big[A t (x-\frac{A t^2}{3m})-E_f\; t\big]}.
\label{4a-9}
\EE\\
\ND
If we set $E_f=0$ and $A=B^3/(2m)$. Then we
have

\BE
\Psi(x,t)
=
Ai\Big[
\Big( \frac{B^3}{\hbar^2} \Big)^{1/3}
\Big(x-\frac{B^3 t^2}{4 m^2} \Big)
\Big]\;
e^{\frac{i}{\hbar} \big[\frac{B^3 t}{2 m }(x-\frac{B^3 t^2}{6 m^2})\big]}.
\label{4a-10}
\EE\\
This is the result derived by Berry and Balaz.
Thus, starting from a static potential $V(x)=A x$, we show that a system with potential energy $V_{nswp}(x,t)=0$ can  supply NSWPs. The shape of the packet   is of the form as  Airy function and moves in a constant acceleration.

We can also explain why Airy packets can accelerate in a free space.
%We note that  a wave packet in free space does not necessarily move in a constant velocity.
The reason is because the motion of an Airy packet is governed not by the total Hamiltonian, but by the Hamiltonian, $H_c(t)$. Referring to Eq. (\ref{3-8}), we have
\BE
H_c(t)
=
\frac{A t}{m} P- A x + \frac{A t^2}{2m}.
\label{4a-11}
\EE\
We see that $H_c(t)$ is not a constant in free space.
And, in fact,  $H_c(t)$ contains a term as $ A x$ which results a constant force $F= A$. It is this force that accelerates the Airy packet.
Thus, we should keep in mind the important role played by the Hamiltonian $H_c(t)$ when analyzing the time evolution of  a quantum packet.
%interesting phenomenon that the dynamics is, in fact, not governed by the total Hamiltonian. The dynamics of motion is, in fact, governed by $H_c(t)$. Thus, an NSWP in a free space need not be in a constant velocity motion.  

%\section{\label{sec:V} Berry and Balazs's NSWPs (2)}

\subsection{\label{subsection:IV-2} Berry and Balazs's NSWPs (2) }

%\textbf{ 4-2. Berry and Balazs's NSWPs (2)}

Eq. (\ref{4a-4}) shows that we may extend from the case of a free space to a more general system to construct NSWPs. We note that we may choose the values for $A$ and $G(t)$ in Eq. (\ref{4a-4}) as

\BA
&&
A - m \ddot{d}(t) =F(t),
\label{4b-1}
\\
&&
G(t) - A  d(t) = 0,
\label{4b-2}
\EA\

\ND
where $F(t)$ is an arbitrary function.
Substituting Eq. (\ref{4b-1}) and (\ref{4b-2}) into (\ref{4a-4}), we obtain

\BE
V_{nswp}(x,t) = - F(t) x.
\label{4b-3}
\EE\\
We then show that a system with a potential energy $V_{nswp}(x,t)= -F(t) x$ can also supply NSWPs. This result was also found by Berry and Balazs in 1979. We note that, in this case,  $V_{nswp}(x,t)$ may be time dependent.
The function $d(t)$ can be solved from Eq.(\ref{4b-1}), and $G(t)$ can be determined from Eq.(\ref{4b-2}). The phase function $\phi(x,t)$
can also be determined from Eqs. (\ref{2-20})-(\ref{2-21}).
We have the result:
\BE
\Psi(x,t)
=
Ai\Big[
\Big( \frac{2 A m}{\hbar^2} \Big)^{1/3}
\Big(x-\frac{E_f}{A}-d(t)\Big)
\Big]\;
e^{i \big(\phi_1(t) x + \phi_0(t)\big)},
\label{4b-4}
\EE\\
\ND
where

\BA
&&
d(t)
= \frac{A t^2}{2m}+\frac{1}{m}\int_0^t\int_0^\tau
F(x)dxd\tau.
\label{4b-5}
\\
&&
\phi_1(t)
=
\frac{A t}{\hbar}+
\frac{1}{\hbar}
\int_0^tF(\tau)d\tau.
\label{4b-6}
\\
&&
\phi_0(t)
=
-\frac{ E_f\; t}{\hbar}
-
\frac{A^2  t^3}{3 m \hbar }
-\frac{1}{2 m \hbar}
\int_0^t\; \Bigg[ \int_0^{\tau} F(\eta) d\eta\;\Bigg]^{2}
d\tau
\nonumber\\
&&
\;\; \;\;\;\;\;\;\;\;\;\;
-\frac{A}{m \hbar }\;
\Bigg[\int_0^t \tau \int_0^{\tau} F(\eta) d\eta
d\tau +\int_0^t \int_0^{\tau} \int_0^\eta  F(\zeta) d\zeta d\eta d\tau
\Bigg].
\label{4b-7}
\EA\\

\ND
If we set $E_f=0$ and $A=B^3/(2m)$, we then reproduce Berry and Balaz's result for the case $V_{nswp}(x,t)= -F(t) x$.
\\
%
%
%We then also have the interesting result, that  such a potential in (\ref{4-3}) can offer NSWPs, as also found by Berry and Balazs in 1979.
%
%
From Eq.(\ref{4a-4}), we see that there is no other way to construct NSWPs with Airy function as the shape function.
We have also derived this result in  Ref. \cite{Lin2}
%
%It had been shown that no other potentials except those described above will support nonspreading Airy packets \cite{Lin2}. 
%\section{\label{sec:VI} Schr\"{o}dinger and Senitzky's NSWPs}

\subsection{\label{subsection:IV-3}  Schr\"{o}dinger and Senitzky's NSWPs }

%\textbf{ 4-3. Schr\"{o}dinger and Senitzky's NSWPs}

We now consider constructing NSWPs in the system of simple harmonic oscillator (SHO). We start from the potential energy as

\BE
V(x)=\frac{1}{2}m\omega^2x^2.
\label{4c-1}
\EE\

This potential energy leads to the well known eigenvalue equation.
\BA
&&
-\frac{\hbar^2}{2m}\partial_x\partial_x \psi_n(x) +\frac{1}{2} m \omega^2 x^2 \psi_n(x) = E_n \psi_n(x),
\label{4c-2}
\\
&&
E_n= (n+\frac{1}{2}) \hbar \omega,.
\label{4c-3}
\EA\\
where $\psi_n(x)$ is the nth eigenfunction of SHO.
This eigenfunction $\psi_n(x)$ will then be used as the shape function for NSWPs.
The corresponding potential energy $V_{nswp}(x,t)$  can be obtained from (\ref{2-18}). This yields

\BA
V_{nswp}(x,t)
&=&
\frac{1}{2} m \omega^2(x-d(t))^2 - m \ddot{d}(t) x +G(t)
\label{4c-4}
\\
&=&
\frac{1}{2} m \omega^2 x^2
- m \Big( \omega^2 d(t) +\ddot{d}(t)\Big) x
+\Big(G(t)+ \frac{1}{2} m \omega^2 d(t)^2 \Big),
\label{4c-5}
\EA\\
To obtain Schr\"odinger 's result, the 
$V_{nswp}(x,t)$ should be time independent. We may then set
\BA
&&
\omega^2 d(t) +\ddot{d}(t)=0,
\label{4c-6}
\\
&&
G(t)+ \frac{1}{2} m \omega^2 d(t)^2=0.
\label{4c-7}
\EA

\ND
This means that
\BE
V_{nswp}(x,t)=\frac{1}{2} m \omega^2 x^2.
\label{4c-8}
\EE\\
This then shows that a simple harmonic oscillator system can offer NSWPs. This was found by Schr\"odinger  in 1926 and Senitzky in 1954.
The functions $d(t)$ and $G(t)$ can be determined from Eqs.(\ref{4c-6}), (\ref{4c-7}). 
We then have the NSWPs described in the following:
\BE
\Psi(x,t)
=\psi_n(x-d(t))
e^{i [\phi_1(t) x + \phi_0(t)]},
\label{4c-9}
\EE\
where
\BA
&&
d(t)
= A \sin(\omega t),
\label{4c-10}
\\
&&
\phi_1(t)
=
\frac{1}{\hbar} m A \omega \cos(\omega t),
\label{4c-11}
\\
&&
\phi_0(t)
=
-\frac{m  \omega x_0^2}{4 \hbar}  \;sin(2\omega
t) -\frac{E_n \; t}{\hbar}.
\label{4c-12}
\EA\
\ND
Above, we set $d(0)=0$, and $\phi_0(0)=0$.
The NSWPs in the SHO system execute a simple harmonic motion.
We have therefore reproduced Schr\"odinger and Senitzky's results.

From Eq. (\ref{4c-5}), we see that we can not arrange $V_{nswp}(x,t)$ into a form as $m\tilde{w}(t)^2 x^2/2$, where $\tilde{w}(t)$ is a time dependent frequency. Thus, a harmonic oscillator system with a time dependent frequency $\tilde{w}(t)$ can not
support a nonspreading wave packet, as stated in Yan's paper
\cite{Yan}.

\section{\label{sec:V}
the generalization to three dimensions }

Our result can easily be generalized to three dimensions.
The same,
we start from a static potential $V(\vec{r})$, then we have the eigenvalue equation:

\BE
-\frac{\hbar^2}{2m}\vec{\nabla} \cdot
\vec{\nabla} f(\vec{r})
+ V(\vec{r})  f(\vec{r})= E_f f(\vec{r}).
\label{5-1}
\EE\\
The corresponding NSWP-construction potential energy $V_{nswp}(\vec{r}, t)$ is of the form as:

\BA
&&
V_{nswp}(\vec{r},t)
=
V\Big( \vec{r}-\vec{d}(t) \Big)-m \vec{r} \cdot \ddot{\vec{d}}(t) + G(t),
\label{5-2}
\\
&&
G(t)
\equiv
-E_f
-\frac{m}{2} \dot{\vec{d}} (t)^2
- \hbar \dot{\phi}_0(t).
\label{5-3}
\EA\\
The wave functions of the NSWPs constructed from $V_{nswp}(\vec{r},t)$ are described as follows
\BA
&&
\Psi(\vec{r},t)=f(\vec{r}-\vec{d}(t))\;e^{ i \phi(\vec{r},t)},
\label{5-4}
\\
&&
\phi(\vec{r},t)=
\frac{m}{\hbar} \dot{\vec{d}}(t) \cdot \vec{r}
 + \phi_0(t),
\label{5-5}
\\
&&
\phi_0(t)
=
-\frac{1}{\hbar•} \int_{0}^{t} 
 \Big(E_f+G(t)+\frac{m}{2} \dot{\vec{d}}(t)^2 \Big) dt.
\label{5-6}
\EA\\
%
%
%And $\phi_0(t)$ is determined from the equation
%
%
%\BE
%\dot{\phi}_0(t)=
%-\frac{1}{\hbar} \Big(E+G(t)+\frac{m}{2} \dot{\vec{d}}(t)^2 \Big).
%\label{6c-4}
%\EE
%
%
We easily check that the $\Psi(\vec{r},t)$
given in Eqs.(\ref{5-4})-(\ref{5-6}) satisfies the Schr\"odinger equation

\BE
i\hbar\frac{\partial \Psi(\vec{r},t)}{\partial t}
=
-\frac{\hbar^2}{2m}\vec{\nabla} \cdot
\vec{\nabla}  \Psi(\vec{r},t)
+ V_{nswp}(\vec{r},t)  \Psi(\vec{r},t).
\label{5-7}
\EE\\

\begin{acknowledgments}
The author is indebted to Prof. Tsin-Fu Jiang for many help and interesting discussions.
\end{acknowledgments}

\newpage

\Large

\end{document}